\documentclass{elsart}

\usepackage{harvard}

\usepackage{graphicx}
\usepackage{epsfig}

\usepackage{amssymb}

\def\url#1{{\ttfamily\def\/{/\discretionary{}{}{}}#1}}

\begin{document}

\begin{frontmatter}
\title{Effect of Expansion and Magnetic Field \\ Configuration on Mass
Entrainment of Jets}


\author[Bama]{A. Rosen\thanksref{ar}}, 
\author[Bama]{P.E. Hardee\thanksref{ph}}

\thanks[ar]{E-mail: rosen@eclipse.astr.ua.edu}
\thanks[ph]{E-mail: hardee@athena.astr.ua.edu}

\address[Bama]{University of Alabama, Department of Physics \& Astronomy \\ Tuscaloosa, AL 35487}

\begin{abstract}

We investigate the growth of jet plus entrained mass in simulations of
supermagnetosonic cylindrical and expanding jets.  The entrained mass
spatially grows in three stages: from an initially slow spatial rate to
a faster rate and finally at a flatter rate.   These stages roughly
coincide with the similar rates of expansion in simulated radio
intensity maps, and also appear related to the growth of the
Kelvin-Helmholtz instability through linear, nonlinear, and saturated
regimes.  In the supermagnetosonic cylindrical jets, we found that a
jet with an embedded primarily toroidal magnetic field is more stable
than a jet with a primarily axial magnetic field.  Also,
pressure-matched expanding jets are more stable and entrain less mass
than cylindrical jets with equivalent inlet conditions.

\end{abstract}

\end{frontmatter}


\vspace{-0.5cm}
\section{Motivation}
\label{motiv}
\vspace{-0.4cm}

Mass entrainment affects the propagation of jets, and at least one
model suggests that its presence may be a distinguishing factor between
the two Fanaroff-Riley morphological types of radio galaxies
\cite{bic94,bic96}.  This work investigates whether there are
observable consequences of jets entraining their surrounding medium and
also determines the extent of the role that the Kelvin-Helmholtz (or
KH) instability has in mass entrainment of magnetized jets.
Acceleration and collimation numerical models, e.g.,
\cite{lov96,kon89,pud97} require that a strong magnetic field is
associated with the jet near its origin.  Jets exit the region near the
origin with magnetic fields aligned primarily along the jet axis,
i.e.\ an axial field, or primarily wrapped around the jet axis, i.e.\ a
toroidal or helical field.  Observations of extended jets, e.g., 4C
32.69, \cite{pot80}; M87, \cite{owe89}; and NGC 6251, \cite{bir93}
indicate that the jet is overpressured with respect to an external
medium and in need of additional confinement that could be provided by
magnetic fields.  Linear analysis of the KH instability suggests that
growth lengths of KH modes are proportional to the product of
magnetosonic Mach number and radius \cite{har97}, suggesting that
expanding jets are more stable.  Here we report on a comparison of
supermagnetosonic cylindrical jets and expanding supermagnetosonic jets
with roughly equivalent inlet conditions.

\vspace{-0.8cm}
\section{\bf Some Details of the Simulations}
\label{simul}
\vspace{-0.4cm}

All of these simulations were performed with a version of ZEUS-3D with
CMoC \cite{cla96} that solves the transverse momentum transport and
magnetic induction equations simultaneously and in a {\it planar split}
fashion.  A second-order accurate MUTCI scheme \cite{van77} is used for
advection.  Two sets of simulations of equilibrium jets have been
performed, which simulate a jet and cocoon well behind the bow shock
and {\it not} a jet propagating into an ambient ISM/ICM/IGM.

The set of supermagnetosonic, initially cylindrical jets includes three
pairs of simulations, each of which has one simulation with a primarily
axial magnetic field and one with a primarily toroidal magnetic field, for 
dense jets with strong fields, dense jets with weak fields, and light 
jets strong fields.  A complete discussion of these simulations is in
\citeasnoun{ros99}.  For the light jets, we set the jet-to-external
density ratio, $\eta$, equal to 0.25 and for the dense jets, $\eta$ =
4.  In the comparison set of expanding supermagnetosonic axial field
jets, there are  three simulations that have conditions at the inlet
similar (but not identical) to some of the simulations in the first
set.  These jets contain primarily axial magnetic fields and the jet
radius expands to 1.5 times the inlet jet radius, $R_0$, at a position
along the jet axis, $z/R_0$ = 60.

The simulations have been completed on a grid that contains some
combination of uniform and ratioed zones along each axis.  In the
uniform portion of the grid, there are 7.5 zones/$R_0$ along the jet
axis and 15 zones/$R_0$ transverse to the jet axis.  The boundary
conditions for all the simulations are outflow everywhere, except at
the jet inlet.  The jet is precessed with a small transverse 
velocity (with an amplitude of $\sim$ 1\% of the axial jet 
velocity), which breaks the symmetry and
should excite the helical modes of the KH instability.  In all of the
simulations, the precession induces a helical twist in the same sense
as that of the magnetic field helicity and helical wavefronts are at
shallow angles to the helically twisted magnetic field lines.

\vspace{-0.85cm}
\section{Results}
\label{results}
\vspace{-0.4cm}

All figures are shown when the simulation has
reached a quasi-steady state out to $z \sim$ 40--50$R_0$.

\vspace{-0.5cm}
\subsection{Mass Entrainment}
\vspace{-0.4cm}

\begin{figure}[h!]
\vspace{7.0cm}
\includegraphics{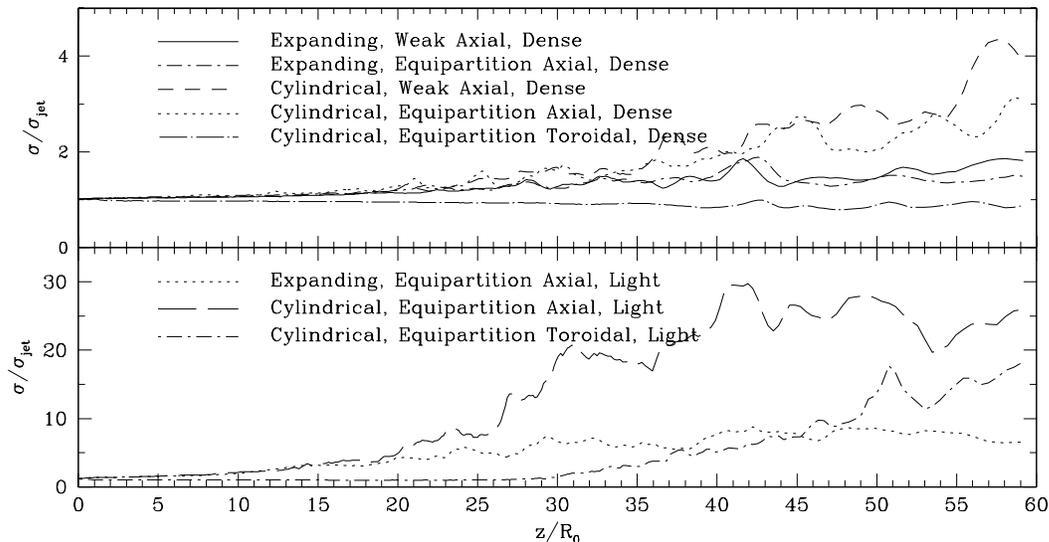}
\caption{Growth of jet plus entrained material with axial position.}
\label{sigfig}
\end{figure}

As a proxy for jet plus entrained mass, we compute the linear mass
density of magnetized mass.  We define the linear mass density,
$\sigma$, at any point along the jet as $\sigma(z) = \int_{\rm A}
f\rho~dx dy$, where A is the cross-sectional area of the computational
domain at $z$, and $f$ is a switch set to 1 if the local magnetic field
is above a threshold value and $f$ = 0 otherwise.  We set the threshold
to $0.04 B_{max}$, where $B_{max}^{~~~2}(z)$ is the expected maximum
magnetic field in the jet at axial position $z$.  In a steady expanding
flow, $B_\phi$ should vary as $R(z)^{-1}$ and $B_z$ should vary as
$R(z)^{-2}$.  Since there is some numerical diffusion of the magnetic
field not associated with the mixing of jet material and unmagnetized
external material, we have found that a threshold based on 4\%
$B_{max}$ is a useful demarkation between ``mixed" (i.e., jet plus
entrained material) and ``unmixed" regions.   In Figure \ref{sigfig},
we show the linear mass density normalized to the initial jet value for
five dense jet (three cylindrical and two expanding) simulations and
three light jet (two cylindrical and one expanding) simulations.


\vspace{-0.5cm}
\subsection{Simulated Intensity Maps}
\vspace{-0.4cm}

\begin{figure}
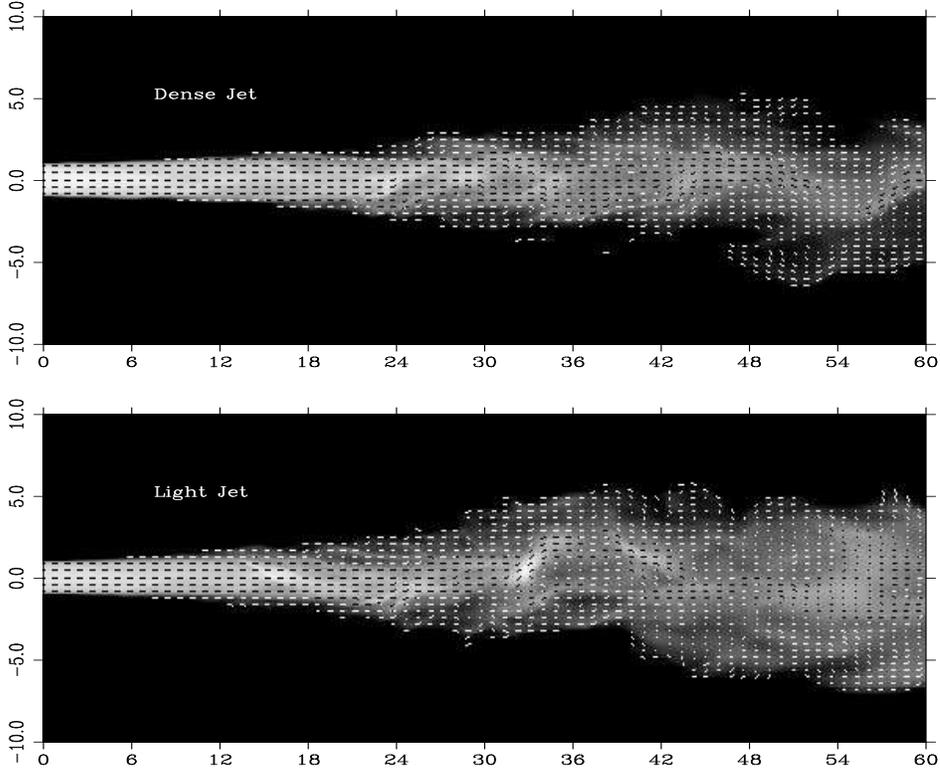

\vspace{10.0cm}
\includegraphics{rosenfig2a.ps}
\includegraphics{rosenfig2b.ps}
\caption{Simulated intensity maps with B-field polarization vectors overlayed.}
\label{radio}
\end{figure}

In Figure \ref{radio}, we display maps of simulated radio intensity,
which are integrations of $p_{th}(B \sin \theta)^{3/2}$, where $\theta$
is the angle between the line-of-sight and the magnetic field, for the
dense and light expanding jet simulations with an equipartition axial
magnetic field.  In order to show the differences in the spine-sheath
structure between the simulations, we have used different grayscales in
each panel of Figure \ref{radio}, in which the range covers 3 orders of
magnitude and the grayscale maximum is 20\% above the actual maximum
intensity in each panel.  The total intensity is overlayed by B-field
polarization vectors that cover regions where the intensity is above
0.001 of the maximum intensity.
  
\vspace{-0.5cm}
\subsection{Example of Structure within Jet}
\vspace{-0.4cm}

\begin{figure}
\vspace{5.3cm}
\includegraphics{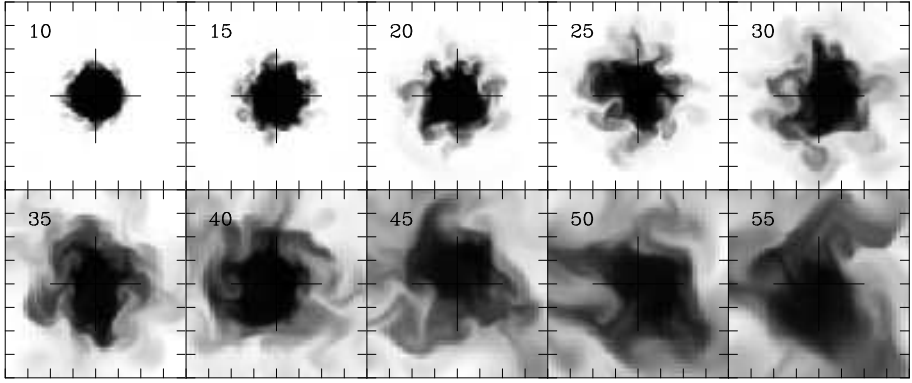}
\caption{Grayscale cross-sections of the axial component of velocity for
the expanding jet simulation with an equipartition axial magnetic field.}
\label{cross}
\end{figure}

The growth of the KH instability and its effect on the jet is
dramatically demonstrated in the grayscale cross-sections of axial
velocity for one simulation (see Figure \ref{cross}).  In this example,
the jet surface has many corrugations, which are characteristic of the
high order KH fluting modes.  In the jet simulations with a primarily
axial magnetic field, the KH instability typically progresses from
higher order modes with small maximum amplitudes of displacement that
dominate close to the inlet to lower order modes with larger distortion
amplitudes farther down the jet.  An example of these low order modes
is the clockwise motion of the jet center about the initial jet axis as
one moves down the jet, suggestive of the helical mode.  A relatively
strong toroidal magnetic field suppresses the growth of the high order
modes, although simulations of jets with this magnetic field
configuration do show evidence for growth of the (low order) helical
mode.

\vspace{-0.5cm}
\section{Discussion}
\label{discussion}
\vspace{-0.4cm}

Spatial growth of magnetized or entrained mass, as measured by the
linear mass density, passes through three stages: an initial slowly
growing stage with a shallow slope, a stage where the growth rate
increases substantially, and a final stage where the linear mass
density does not increase.  The third stage is most noticeable in the
(axial field) light jet simulations, which develop more quickly than
dense jet simulations.  For example, in both axial field simulations in
the bottom panel of Figure \ref{sigfig}, each stage spans roughly
one-third of the simulated jet axis in the light jet with an axial
field.  In jets with a moderate magnetic field strength, either a
primarily toroidal field or jet expansion can reduce the mass entrained
by the jet.   The expanding jet simulations have an isothermal
pressure-matched external medium, which falls off in density.  It is
this decrease in density that is responsible for the smaller amount of
mass entrained in the expanding jet simulations (see the comparison
with the cylindrical equivalents in Figure \ref{sigfig}).  From a
comparison of positions where the transitions between regions of slow,
fast, and no mass entrainment occur in the different sets of
simulations, the expanding supermagnetosonic jets are more stable than
their cylindrical equivalents.

From a comparison of our results with predictions from a linear
stability analysis (for growth lengths of modes), we confirm that the
three stages of growth of the jet plus entrained mass match the stages
of growth for the KH instability as it progresses through linear,
nonlinear, and saturated stages \cite{ros00}.  The linear stage is
associated with the progression from high order modes to low order
modes, the nonlinear with development of large amplitude low order
modes and saturation with the maximum amplitude of the low order
modes.  This progression from high order modes to low order modes is
displayed in cross-sections of axial velocity (e.g., Figure
\ref{cross}).  By the axial position where the flow is dominated by the
elliptical and helical modes, the magnetized ``core" of the jet begins
to disrupt and is no longer noticeable in the simulated intensity
images (e.g., Figure \ref{radio}).  Additionally, the width of the
magnetized mass as it appears in simulated intensity images grows in
three similar stages. Thus, we do find an observable consequence of
mass entrainment.  This progression in the expansion of the jets in the
intensity images is most noticeable in the light jet simulations, where
the region of constant apparent width also has a more uniform intensity
than any portion of the dense jets.  This uniform appearance suggests a
greater mixing of jet and external material in the light jets.


\end{document}